\documentclass[prl,twocolumn,superscriptaddress,showpacs,preprintnumbers,amsmath,amssymb]{revtex4}

\usepackage{graphicx}
\usepackage{dcolumn}
\usepackage{bm}

\begin{document}

\preprint{accepted for publication in Physical Review Letters}

\title{Gyroid Phase in Nuclear Pasta}

\author{Ken'ichiro Nakazato}
 \email{nakazato@heap.phys.waseda.ac.jp}
 \affiliation{Department of Astronomy, Kyoto University, Kita-shirakawa Oiwake-cho, Sakyo, Kyoto 606-8502, Japan
}%

\author{Kazuhiro Oyamatsu}
 \affiliation{Department of Library and Information Science, Aichi Shukutoku University, Nagakute-Katahira 9, Nagakute, Aichi 480-1197, Japan
}%

\author{Shoichi Yamada}
 \altaffiliation[Also at ]{Advanced Research Institute for Science \& Engineering, Waseda University, 3-4-1 Okubo, Shinjuku, Tokyo 169-8555, Japan}
 \affiliation{Department of Physics, Waseda University, 3-4-1 Okubo, Shinjuku, Tokyo 169-8555, Japan
}%

\date{\today}

\begin{abstract}
Nuclear matter is considered to be inhomogeneous at subnuclear densities that are realized in supernova cores and neutron star crusts, and the structures of nuclear matter change from spheres to cylinders, slabs, cylindrical holes and spherical holes as the density increases. In this letter, we discuss other possible structures, that is, gyroid and double-diamond morphologies, which are periodic bicontinuous structures discovered in a block copolymer. Utilizing the compressible liquid drop model, we evaluate their surface and Coulomb energies and show that there is a chance of gyroid appearance near the transition point from a cylinder to a slab. This interesting analogy between nuclear and polymer systems is not merely qualitative. The volume fraction at the phase transition is also similar for the two systems. Although the five shapes listed initially have been long thought to be the only major constituents of so-called nuclear pasta at subnuclear densities, our findings imply that this may not be the case and suggest that more detailed studies on nuclear pasta including the gyroid phase are needed.
\end{abstract}


\maketitle

Nature contains numerous examples of materials with various morphologies at different levels of hierarchy. Nuclear matter is no exception. Terrestrial atomic nuclei are usually almost spherical. About a quarter of a century ago, however, it was shown that nuclei deform at subnuclear densities from spheres (SP) to cylinders (C), slabs (S), cylindrical holes (CH) and spherical holes (SH) as the density increases to $\sim10^{14}$g/cm$^3$, the density of the uniform nuclear matter~\cite{raven83, hashi84}. Owing to the similarity of these shapes to meat balls, spaghetti, lasagna, macaroni and Swiss cheese, respectively, these structures are called nuclear pasta. Nuclei with the pasta structures are important because they are thought to actually exist inside the core of supernovae and the crust of neutron stars and have an impact on astrophysical phenomena such as supernova explosions, proto-neutron star cooling, pulsar glitches and so forth \cite{nbsnd05, moti97}. Assuming that only the five nuclear shapes listed above exist, many detailed studies have been perfomed since this discovery; for reviews, see Refs.~\cite{petrav95, nbsnd05}.

Similar structure transitions are known to occur in nanostructures of block copolymers and, more interestingly, several complex structures have been discovered experimentally \cite{bafre99}. Some of them have periodic bicontinuous structures such as the so-called gyroid (G) and double-diamond (D) morphologies, and are likely to appear between the C and S phases. In this letter, we investigate the possible appearance of G and D morphologies in nuclear pasta by employing the compressible liquid drop model (CLDM), which is a phenomenological model also used in earlier studies \cite{raven83, hashi84, oyak84}. In this approach, nuclei are approximated as charged liquid drops and the nuclear shape is determined by minimizing the total energy density, in which the surface tension competes with the Coulomb repulsion. It is also noted that the nuclear shape is a function of the volume fraction of the nucleus in a unit cell, $u$, similarly to the case of polymer shapes \cite{bafre99}. We then find that there is a good chance that the G morphology appears near the transition point from the C phase to the S phase. Furthermore, the volume fraction of nuclei at this point is $u\sim0.35$, which is very close to the value obtained for the polymer system, suggesting commom underlying physical laws between the two systems. Because the nuclear pasta phases and the gyroid phase in the block copolymer have separately attracted many researchers' interest, our findings may offer a new field of interest.

\begin{figure}
\begin{center}
\includegraphics{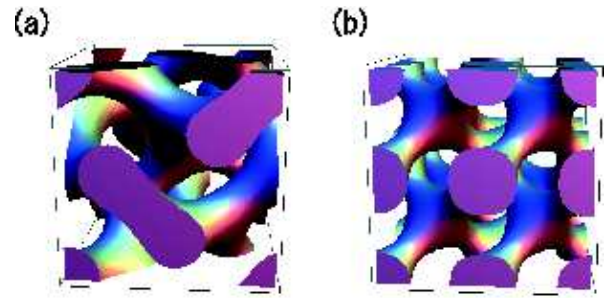}
\caption{Bird's-eye views of unit cubes of (a) gyroid and (b) double-diamond, in which bicontinuous minimum surfaces are shown for volume fraction $u=0.35$.}
\label{kaiju}
\end{center}
\end{figure}
\textbf{Setup.---}The bicontinuous structures observed in block copolymers are thought to have a periodic minimum surface \cite{thomas88}, which is a stationary surface for variations of surface area with a fixed volume fraction. Although the D morphology was initially considered to be the most likely structure, the G morphology is often thought to be a more probable structure. In this letter, we study the double network structures of G and D minimum surfaces as new types of nuclear pasta. 

It is known that these bicontinuous structures can be closely approximated by the following level surfaces:
\begin{eqnarray}
f(x, y, z) & = & \sin \frac{2 \pi x}{a} \cos \frac{2 \pi y}{a} + \sin \frac{2 \pi y}{a} \cos \frac{2 \pi z}{a} \nonumber \\
 & & + \sin \frac{2 \pi z}{a} \cos \frac{2 \pi x}{a} = \pm k,
\label{gyroid}
\end{eqnarray}
for G and
\begin{eqnarray}
f(x, y, z) & = & \cos \frac{2 \pi (x-y)}{a} \cos \frac{2 \pi z}{a} \nonumber \\
 & & + \sin \frac{2 \pi (x+y)}{a} \sin \frac{2 \pi z}{a} = \pm k,
\label{dd}
\end{eqnarray}
for D \cite{schgom02}, where $(x, y, z)$ are the spatial coordinates and $a$ is the size of the unit cube. Nucleons are assumed to reside in the region that satisfies $|f(x, y, z)| > k$, which is called the ``nucleus''. $k$ is a positive parameter that specifies the volume fraction, $u$, of the nucleus in the unit cube. $k \to 0$ corresponds to $u \to 1$ and $u$ is a monotonically decreasing function of $k$. Note that some neutrons may drip out of the nucleus at high densities in the neutron-rich case. The shapes of the ``nuclei'' are illustrated in Fig.~\ref{kaiju}.

Equations~(\ref{gyroid}) and (\ref{dd}) are no longer good approximations for very small values of $u$ (and hence the corresponding values of $k$): $u<0.0354$ for G and $u<0.161$ for D, since the resultant configurations are not bicontinuous but pinched off. Although this poses no problem for the following analysis, we do not consider these configurations. On the other hand, the hole structures of the gyroid and double-diamond, for which nucleons reside in the region satisfying $|f(x, y, z)| <k$, are taken into account. Again the configurations are bicontinuous only for $u<0.965$ for the gyroid hole (GH) and $u<0.839$ for the double-diamond hole (DH).

In our model, the volume of the unit cell is $a^3$ and, as already stated, the nucleus occupies the fraction $u$. The number density of nucleons, $n^\mathrm{in}$, and the proton fraction, $x^\mathrm{in}$, inside the nucleus as well as the number density of dripped neutrons outside the nucleus, $n^\mathrm{out}$, are set to be constant. Electrons are uniformly distributed in the cell and their number density is $ux^\mathrm{in}n^\mathrm{in}$, obtained from the charge neutrality.

\textbf{Analysis.---}Our analysis is an improvement of the previous CLDM studies \cite{raven83, hashi84, oyak84}. As shown below, it is mainly based on a geometrical argument and is insensitive to nuclear interaction models. We write the total energy in the unit cell, $W$, as
\begin{equation}
W = W_b + W_s + W_C,
\label{totale}
\end{equation}
where $W_b$, $W_s$ and $W_C$ are the bulk energy both inside and outside the nucleus including the kinetic energy of electrons, the surface energy and the Coulomb energy, respectively. The use of Eq.~(\ref{totale}) is motivated by the semi-empirical mass formula for terrestrial nuclei. The bulk energy includes not only the volume term but also the symmetry term. Since $W_b$ is proportional to the volume, we rewrite it as
\begin{equation}
W_b = w_b(u, x^\mathrm{in}, n^\mathrm{in}, n^\mathrm{out}) a^3,
\label{bulke}
\end{equation}
where $w_b(u, x^\mathrm{in}, n^\mathrm{in}, n^\mathrm{out})$ is the average energy density.

We assume that the surface energy is proportional to the surface area and, hence, depends on the shape of the nucleus. Note, on the other hand, that $W_b$ is independent of the shape. We rewrite the surface energy in the unit cell as
\begin{equation}
W_s = \sigma(x^\mathrm{in}, n^\mathrm{in}, n^\mathrm{out}) g(u, \mathrm{shape}) a^2,
\label{surfe}
\end{equation}
where $\sigma(x^\mathrm{in}, n^\mathrm{in}, n^\mathrm{out})$ is the surface tension, $g(u, \mathrm{shape})$ is the relative surface area and $\mathrm{shape}=\mathrm{SP}$, C, S, CH, SH, G, D, GH or DH.

The Coulomb energy of the unit cell can be expressed similarly as
\begin{equation}
W_C = \left(e x^\mathrm{in} n^\mathrm{in} \right)^2 w_C(u, \mathrm{shape}) a^5,
\label{couler}
\end{equation}
where $e$ is the elementary charge and $w_C(u, \mathrm{shape})$ is the relative Coulomb energy, which depends on the fraction $u$ as well as on the nuclear shape, and is obtained by numerically solving the Poisson equation for the Coulomb potential by a discrete Fourier transform. Since the Coulomb energy is proportional to the product of two charges divided by their separation and because the total charge in the unit cell is proportional to $a^3$ (Note that the charge density is fixed in our model.), the Coulomb term, $W_C$, in the energy expression is proportional to $a^5$.

Substituting Eqs.~(\ref{bulke})-(\ref{couler}) into Eq.~(\ref{totale}), we minimize the total energy density. This consists of two steps. Firstly, minimization with respect to the size of the unit cube, $a$, is performed by a conventional derivation:
\begin{eqnarray}
\frac{\partial}{\partial a}\left( \frac{W}{a^3} \right) & = & - \frac{\sigma(x^\mathrm{in}, n^\mathrm{in}, n^\mathrm{out}) g(u, \mathrm{shape})}{a^2}  \nonumber \\
 & & + 2 \left(e x^\mathrm{in} n^\mathrm{in} \right)^2 w_C(u, \mathrm{shape}) a  \nonumber \\
 & = & 0.
\label{minitotaled}
\end{eqnarray}
This result is simply the well-known condition for size equilibrium, $W_s = 2W_C$. Eliminating $a$, we obtain
\begin{eqnarray}
\left( \frac{W}{a^3} \right) & = & w_b(u, x^\mathrm{in}, n^\mathrm{in}, n^\mathrm{out}) + \nonumber \\
 & & \frac{3}{\sqrt[3]{4}} \left[ e x^\mathrm{in} n^\mathrm{in} \sigma(x^\mathrm{in}, n^\mathrm{in}, n^\mathrm{out}) \right]^{2/3} \nonumber \\
 & & \times g(u, \mathrm{shape})^{2/3} w_C(u, \mathrm{shape})^{1/3}.
\label{totaledagain}
\end{eqnarray}

\begin{widetext}

\begin{figure}
\begin{center}
\includegraphics{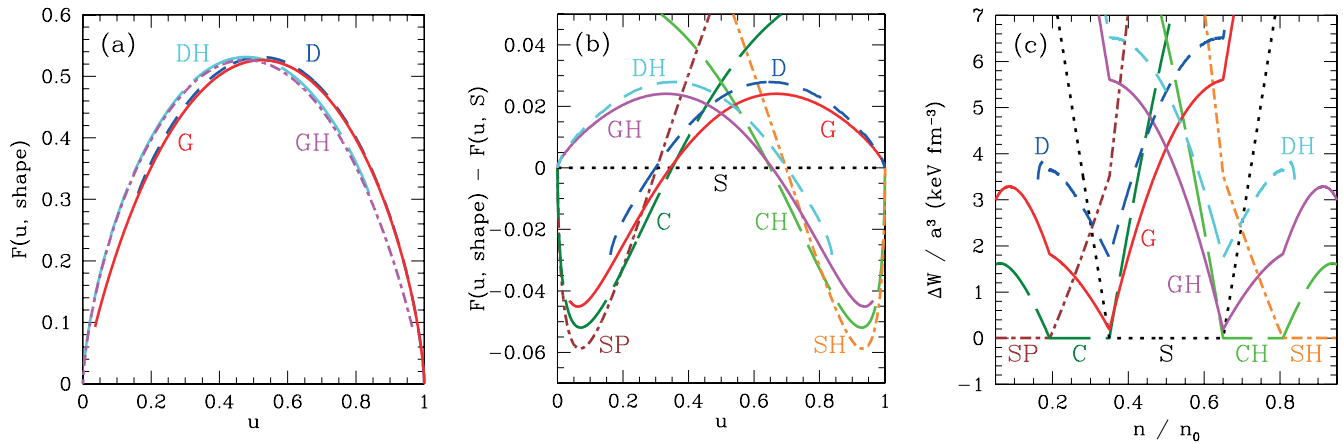}
\caption{(a) Relative energy densities, $F(u, \mathrm{shape})$, as a function of $u$ for the periodic bicontinuous morphologies. (b) Differences between the relative energy densities of various phases and that of the slab phase. (c) Differences between the average energy densities and that of the most stable phase for supernova matter. In all panels, the following notation is used: gyroid (G), double-diamond (D), gyroid hole (GH), double-diamond hole (DH), sphere (SP), cylinder (C), slab (S), cylindrical hole (CH) and spherical hole (SH).}
\vspace{-12mm}
\label{resultfig}
\end{center}
\end{figure}

\end{widetext}

Secondly, we minimize Eq.~(\ref{totaledagain}) with respect to the shape for a given $u$. This is performed by simply comparing the energies for different shapes and finding the shape that gives the lowest energy. Note that in Eq.~(\ref{totaledagain}) the shape dependence is entirely encapsulated in the relative energy density, $F(u, \mathrm{shape})$, defined as
\begin{equation}
F(u, \mathrm{shape}) = g(u, \mathrm{shape})^{2/3} w_C(u, \mathrm{shape})^{1/3}.
\label{fighty}
\end{equation}
As a result, the nuclear shape does not depend on the nuclear interaction models, which are encoded in the average energy density, $w_b(u, x^\mathrm{in}, n^\mathrm{in}, n^\mathrm{out})$, and the surface tension, $\sigma(x^\mathrm{in}, n^\mathrm{in}, n^\mathrm{out})$.

For the conventional pasta phases, the relative energy densities are expressed as follows:
\begin{subequations}
\begin{multline}
F(u, \mathrm{SP}) = \\
\shoveleft{ \left( 36 \pi u^2 \right)^{2/9} \left[ \frac{\sqrt[3]{9 \pi}
\left( 2u^{5/3} - 3u^2 + u^{8/3} \right)}{5 \sqrt[3]{2}} + c_\mathrm{bcc}u^2 \right]^{1/3},}
\label{spheref}
\end{multline}
\vspace{-8mm}
\begin{multline}
F(u, \mathrm{C}) = \left( 4 \pi u \right)^{1/3} \left[ \frac{u^2}{2} \left( u - 1 - \log u \right) + 
c_\mathrm{hex}u^2 \right]^{1/3},
\label{clindf}
\end{multline}
\vspace{-10mm}
\begin{equation}
F(u, \mathrm{S}) = \left( \frac{2 \pi}{3} \right)^{1/3} u^{2/3} (1-u)^{2/3}.
\label{sladf}
\end{equation}
\label{pastaf}
\end{subequations}
The numerically determined coefficients, $c_\mathrm{bcc}=6.5620\times10^{-3}$ and $c_\mathrm{hex}=1.2475\times10^{-3}$, are corrections to the Wigner-Seitz (WS) approximation, and the subscripts ``bcc'' and ``hex'' represent the body-centered cubic and hexagonal lattices, respectively, which are the most stable alignments for each nuclear shape \cite{oyak84}. For the hole nuclei, the following relations hold: $F(u, \mathrm{CH}) = F(1-u, \mathrm{C})$ and $F(u, \mathrm{SH}) = F(1-u, \mathrm{SP})$.

\textbf{Results.---}In Fig.~\ref{resultfig}(a), we show $F(u, \mathrm{shape})$ as functions of $u$ for the periodic bicontinuous shapes. It is found that the G phase always has a lower energy than the D phase. The curves of G (D) and GH (DH) are reflections of each other in the vertical line $u=1/2$ for the following reason. The hole morphology with a fraction of $u$ has the same surface area as the normal morphology with a fraction of $1-u$. Hence, relations such as $g(u, \mathrm{GH})=g(1-u, \mathrm{G})$ hold. In a similar way, it can be easily understood that the hole morphology has a charge density and Coulomb potential with opposite signs to those of their normal counterparts. Since the Coulomb energy is a product of the charge density and Coulomb potential, relations such as $w_C(u, \mathrm{GH})=w_C(1-u, \mathrm{G})$ are satisfied. As a result, the most stable bicontinuous phase is G for $u\leq0.5$ and GH for $u\geq0.5$. Note that the curves for G and D converge to 0 at the point $u=1$, which corresponds to uniform matter. In Fig.~\ref{resultfig}(b), we show $F(u, \mathrm{shape})-F(u, \mathrm{S})$ for all the shapes in nuclear pasta. We find that the relative energy density of the G phase becomes very close to those of the C and S phases at the transition point from the C phase to the S phase ($u=0.35$). The same is true of the transition point from the S phase to the CH phase ($u=0.65$).

To evaluate the energy difference quantitatively, we must fix the parameters. Bearing in mind the application to the core of supernovae, we adopt a simple estimate using the incompressible liquid drop model employed in Ref.~\cite{nbsnd05}. We set $n^\mathrm{out}=0$ and $n^\mathrm{in}=n_0$, where $n_0=0.165$~fm$^{-3}$ is the saturation density, and the volume fraction is given as $u=n/n_0$, where $n$ is the average number density of nucleons. The surface tension is assumed to be $\sigma=0.73$~MeV~fm$^{-2}$, the value that reproduces the properties of isolated finite nuclei in the limit of $u \to 0$. The proton fraction is set to $x^\mathrm{in}=0.3$.

The difference between the average energy density for each shape and that of the most stable phase, $\Delta W / a^3$, for each shape is computed using these parameters and is shown in Fig.~\ref{resultfig}(c). It can be seen that $\Delta W / a^3$ for the G phase becomes only $\sim0.2$~keV~fm$^{-3}$ at $n = 0.35n_0$ ($u=0.35$). This is comparable to the hexagonal lattice correction to the WS approximation for cylindrical nuclei at $u=0.35$. For the matter inside the crust of neutron stars, the minimum of $\Delta W / a^3$ will be even smaller by a factor of a few than that for the matter in supernovae. Note that the neutron drip is not negligible for matter in a neutron star crust and, as a result, the transition density will be somewhat larger than $n=0.35n_0$.

\textbf{Discussion.---}The G phase is not the most stable phase for any $n$ (or $u$). Considering the tiny difference in energy densities, however, it is possible for the G phase to exist as a metastable state. In fact, recent studies on the dynamics of pasta phases by quantum molecular dynamics (QMD) show that the phase transition from C to S is a dynamical process \cite{wtnb05} and that intermediate phases, which are different from any of the known pasta phases, may emerge between the C and S phases as well as between the S and CH phases \cite{sndy08}. These results indicate the appearance of metastable states, which may include the G phase. Since the dynamical stability of the G phase cannot be assessed by our model, our argument is speculative, but this is an interesting issue to be pursued further in the future.

The coexistence of the G and other phases is even more likely at finite temperatures. If this occurs, it should have an impact on the thermodynamical properties of nuclear matter such as the equation of state. In fact, the minimum value of $\Delta W / a^3$ for the G phase is $\sim0.2$~keV~fm$^{-3}$, or $\sim3$~keV per nucleon, at $n = 0.35n_0$, which is much smaller than the temperatures of several MeV in supernova cores. Hence, the G phase will almost certainly exist as a thermal fluctuation. The surface temperatures of some neutron stars are several keV. Since $\Delta W / a^3$ for matter in a neutron star crust is smaller than that for supernova matter, it is also possible that the G phase will appear in neutron star crusts. Note that the shell effect, which is also a constituent of the semi-empirical mass formula but has been estimated to be minor ($\lesssim 2$~keV per nucleon) for pasta nuclei \cite{oyak94} and thus was omitted in our model, might also be helpful for the G phase appearance.

In Ref.~\cite{matsuz06}, the difference between the energy density of the D and S phases was calculated and was found to be much larger than that obtained in this paper. This discrepancy originates from the fact that in Ref.~\cite{matsuz06} the total energy was not minimized with respect to $a$. Since the energy differences between phases are small for nuclear pasta, the minimization is crucial.

We have already pointed out an interesting quantitative similarity between nuclear and polymer systems. The G phase in the block copolymer is often observed in a narrow range of the volume fraction near $\sim0.35$ \cite{laurer97}. This is exactly where $F(u, \mathrm{G})$ becomes very close to $F(u, \mathrm{C})$ and $F(u, \mathrm{S})$ in the CLDM. Since Coulomb screening cannot be neglected in the polymer system, a simple analogy hardly seems applicable. Our results, however, appear to suggest some common underlying physical principles.~\cite{ohta06}

In this letter we have demonstrated, contrary to conventional wisdom, that there is a good chance of the G phase appearance. There are some issues remaining unresolved in our model, however. Although we assumed the shape of the bicontinuous surface a priori, the energy density may be further decreased by varying the shape. The finite thickness of the surface, which is not taken into account in the CLDM, is also expected to be important. Therefore, further theoretical investigations with more advanced treatments of the surface are necessary. The dynamical stability of the G phase, finite-temperature corrections to the free energy and the possible implications for astrophysical phenomena (e.g., supernova explosions, proto-neutron star cooling and pulsar glitches) will be also interesting issues. Although there are still some controversies, even regarding the appearance of the conventional pasta phases, we strongly urge that the new type of nuclear pasta is included in these detailed investigations.

\begin{acknowledgments}
This work was partially supported by JSPS and Grants-in-Aid from MEXT, Japan through No.~20105004, No.~21540281.
\end{acknowledgments}

\bibliographystyle{apsrev}
\bibliography{apssamp}

\begin{thebibliography}{10}
\expandafter\ifx\csname natexlab\endcsname\relax\def\natexlab#1{#1}\fi
\expandafter\ifx\csname bibnamefont\endcsname\relax
  \def\bibnamefont#1{#1}\fi
\expandafter\ifx\csname bibfnamefont\endcsname\relax
  \def\bibfnamefont#1{#1}\fi
\expandafter\ifx\csname citenamefont\endcsname\relax
  \def\citenamefont#1{#1}\fi
\expandafter\ifx\csname url\endcsname\relax
  \def\url#1{\texttt{#1}}\fi
\expandafter\ifx\csname urlprefix\endcsname\relax\def\urlprefix{URL }\fi
\providecommand{\bibinfo}[2]{#2}
\providecommand{\eprint}[2][]{\url{#2}}

\bibitem[{\citenamefont{Ravenhall et~al.}(1983)\citenamefont{Ravenhall,
  Pethick, and Wilson}}]{raven83}
\bibinfo{author}{\bibfnamefont{D.~G.} \bibnamefont{Ravenhall}},
  \bibinfo{author}{\bibfnamefont{C.~J.} \bibnamefont{Pethick}},
  \bibnamefont{and} \bibinfo{author}{\bibfnamefont{J.~R.}
  \bibnamefont{Wilson}}, \bibinfo{journal}{Phys.\ Rev.\ Lett.}
  \textbf{\bibinfo{volume}{50}}, \bibinfo{pages}{2066} (\bibinfo{year}{1983}).

\bibitem[{\citenamefont{Hashimoto et~al.}(1984)\citenamefont{Hashimoto, Seki,
  and Yamada}}]{hashi84}
\bibinfo{author}{\bibfnamefont{M.}~\bibnamefont{Hashimoto}},
  \bibinfo{author}{\bibfnamefont{H.}~\bibnamefont{Seki}}, \bibnamefont{and}
  \bibinfo{author}{\bibfnamefont{M.}~\bibnamefont{Yamada}},
  \bibinfo{journal}{Prog.\ Theor.\ Phys.} \textbf{\bibinfo{volume}{71}},
  \bibinfo{pages}{320} (\bibinfo{year}{1984}).

\bibitem[{\citenamefont{Watanabe and Sonoda}(2007)}]{nbsnd05}
\bibinfo{author}{\bibfnamefont{G.}~\bibnamefont{Watanabe}} \bibnamefont{and}
  \bibinfo{author}{\bibfnamefont{H.}~\bibnamefont{Sonoda}},
  in \emph{\bibinfo{title}{Soft Condensed Matter: New Research}}, edited by \bibinfo{editor}{\bibfnamefont{K.~I.}~\bibnamefont{Dillon}}
  (\bibinfo{publisher}{New York: Nova Science Publishers}, \bibinfo{year}{2007}).

\bibitem[{\citenamefont{Mochizuki et~al.}(1997)\citenamefont{Mochizuki,
  Oyamatsu, and Izuyama}}]{moti97}
\bibinfo{author}{\bibfnamefont{Y.~S.} \bibnamefont{Mochizuki}},
  \bibinfo{author}{\bibfnamefont{K.}~\bibnamefont{Oyamatsu}}, \bibnamefont{and}
  \bibinfo{author}{\bibfnamefont{T.}~\bibnamefont{Izuyama}},
  \bibinfo{journal}{Astrophys.\ J.} \textbf{\bibinfo{volume}{489}},
  \bibinfo{pages}{848} (\bibinfo{year}{1997}).

\bibitem[{\citenamefont{Pethick and Ravenhall}(1995)}]{petrav95}
\bibinfo{author}{\bibfnamefont{C.~J.} \bibnamefont{Pethick}} \bibnamefont{and}
  \bibinfo{author}{\bibfnamefont{D.~G.} \bibnamefont{Ravenhall}},
  \bibinfo{journal}{Annu.\ Rev.\ Nucl.\ Part.\ Sci.}
  \textbf{\bibinfo{volume}{45}}, \bibinfo{pages}{429} (\bibinfo{year}{1995}).

\bibitem[{\citenamefont{Bates and Fredrickson}(1999)}]{bafre99}
\bibinfo{author}{\bibfnamefont{F.~S.} \bibnamefont{Bates}} \bibnamefont{and}
  \bibinfo{author}{\bibfnamefont{G.~H.} \bibnamefont{Fredrickson}},
  \bibinfo{journal}{Phys.\ Today} \textbf{\bibinfo{volume}{52}}, No.~\bibinfo{number}{2}, \bibinfo{pages}{32} (\bibinfo{year}{1999}).

\bibitem[{\citenamefont{Oyamatsu et~al.}(1984)\citenamefont{Oyamatsu,
  Hashimoto, and Yamada}}]{oyak84}
\bibinfo{author}{\bibfnamefont{K.}~\bibnamefont{Oyamatsu}},
  \bibinfo{author}{\bibfnamefont{M.}~\bibnamefont{Hashimoto}},
  \bibnamefont{and} \bibinfo{author}{\bibfnamefont{M.}~\bibnamefont{Yamada}},
  \bibinfo{journal}{Prog.\ Theor.\ Phys.} \textbf{\bibinfo{volume}{72}},
  \bibinfo{pages}{373} (\bibinfo{year}{1984}).

\bibitem[{\citenamefont{Thomas et~al.}(1988)\citenamefont{Thomas, Anderson,
  Henkee, and Hoffman}}]{thomas88}
\bibinfo{author}{\bibfnamefont{E.~L.} \bibnamefont{Thomas}} \bibnamefont{{\it et~al.}},
  \bibinfo{journal}{Nature (London)} \textbf{\bibinfo{volume}{334}},
  \bibinfo{pages}{598} (\bibinfo{year}{1988}).

\bibitem[{\citenamefont{Schwarz and Gompper}(2002)}]{schgom02}
\bibinfo{author}{\bibfnamefont{U.}~\bibnamefont{Schwarz}} \bibnamefont{and}
  \bibinfo{author}{\bibfnamefont{G.}~\bibnamefont{Gompper}},
  in \emph{\bibinfo{title}{Morphology of Condensed Matter: Physics and Geometry of Spatially Complex Systems}}, edited by \bibinfo{editor}{\bibfnamefont{K.}~\bibnamefont{Mecke}} \bibnamefont{and} \bibinfo{editor}{\bibfnamefont{D.}~\bibnamefont{Stoyan}}
  (\bibinfo{publisher}{Berlin: Springer}, \bibinfo{year}{2002}).

\bibitem[{\citenamefont{Watanabe et~al.}(2005)\citenamefont{Watanabe, Maruyama,
  Sato, Yasuoka, and Ebisuzaki}}]{wtnb05}
\bibinfo{author}{\bibfnamefont{G.}~\bibnamefont{Watanabe}} \bibnamefont{{\it et~al.}},
  \bibinfo{journal}{Phys.\ Rev.\ Lett.} \textbf{\bibinfo{volume}{94}},
  \bibinfo{pages}{031101} (\bibinfo{year}{2005}).

\bibitem[{\citenamefont{Sonoda et~al.}(2008)\citenamefont{Sonoda, Watanabe,
  Sato, Yasuoka, and Ebisuzaki}}]{sndy08}
\bibinfo{author}{\bibfnamefont{H.}~\bibnamefont{Sonoda}} \bibnamefont{{\it et~al.}},
  \bibinfo{journal}{Phys.\ Rev.\ C} \textbf{\bibinfo{volume}{77}},
  \bibinfo{pages}{035806} (\bibinfo{year}{2008}).

\bibitem[{\citenamefont{Oyamatsu and Yamada}(1994)}]{oyak94}
\bibinfo{author}{\bibfnamefont{K.}~\bibnamefont{Oyamatsu}} \bibnamefont{and}
  \bibinfo{author}{\bibfnamefont{M.}~\bibnamefont{Yamada}},
  \bibinfo{journal}{Nucl.\ Phys.\ A} \textbf{\bibinfo{volume}{578}},
  \bibinfo{pages}{181} (\bibinfo{year}{1994}).

\bibitem[{\citenamefont{Matsuzaki}(2006)}]{matsuz06}
\bibinfo{author}{\bibfnamefont{M.}~\bibnamefont{Matsuzaki}},
  \bibinfo{journal}{Phys.\ Rev.\ C} \textbf{\bibinfo{volume}{73}},
  \bibinfo{pages}{028801} (\bibinfo{year}{2006}).

\bibitem[{\citenamefont{Laurer et~al.}(1997)}]{laurer97}
\bibinfo{author}{\bibfnamefont{J.~H.} \bibnamefont{Laurer}} \bibnamefont{{\it et~al.}},
  \bibinfo{journal}{Macromolecules} \textbf{\bibinfo{volume}{30}},
  \bibinfo{pages}{3938} (\bibinfo{year}{1997}).

\bibitem[{\citenamefont{Ohta}(2006)}]{ohta06}
\bibinfo{author}{\bibfnamefont{T.}~\bibnamefont{Ohta}},
  \bibinfo{journal}{Prog.\ Theor.\ Phys.\ Suppl.}
  \textbf{\bibinfo{volume}{164}}, \bibinfo{pages}{203} (\bibinfo{year}{2006}).

\end{thebibliography}

\end{document}